\documentstyle [12pt]{article}
\begin{document}
\setcounter{page}1
\setcounter{footnote}0

\begin{flushright}
\large\bf
KIPT E96-1
\end{flushright}

\begin{center}
{\large \bf
National Science Center\\
"Kharkov Institute of Physics and Technology" \\
\vspace{5cm}
M.I.Ayzatsky\footnote{ M.I.Ayzatsky (N.I.Aizatsky)\\
National Science Center
"Kharkov Institute of Physics and Technology"\\
Akademicheskaya 1,
Kharkov, 310108, Ukraine\\
e-mail:aizatsky@nik.kharkov.ua}\\
\vspace{2cm}
ON TWO-CAVITY COUPLING\footnote{The paper was presented to PAC95
and will be published in ZhTF(1996)}\\
\vspace{1.5cm}
E-Preprint\\
\vspace{6cm}
Kharkov --- 1996}
\end{center}
\newpage

\begin{abstract}
This work presents research results on a novel analytical model
of electromagnetic system coupling through small size holes. The
key problem regarding the coupling of two cavities through an
aperture in separating screen of finite thickness without making
assumption on smallness of any parameters is considered. We are
the first to calculate on the base of rigorous electromagnetic
approach the coupling coefficients of the cylindrical cavities within
the limit of small aperture and infinitely thin separating screen. The
numeric results of electromagnetic characteristic dependencies that
have been impossible to perform on the base of previous models are
given.
\end{abstract}

\section{Introduction}

The problem of electromagnetic coupling has been in
the focus of scientific attention for over 40 years. The approach of
tackling this problem with the use of the concepts of equivalent
electric and magnetic dipole moments, suggested in \cite{r1,r2}, proved to
be fruitful. On its base various electromagnetic characteristics of
interacting objects have been studied (see \cite{r3}-\cite{r10}) and
literature cited therein). The key element of this approach is employment of
the static analysis used for determination the fields in the immediate vicinity
of the hole. Clearly, this procedure is valid only if the hole dimensions
are small compared to the wave length. Besides, the apertures have to be
placed at a remote distance from the borders of the electromagnetic systems
being considered. This notwithstanding, the developed methods allowed not
only to calculate the number of important characteristics, but formulate (or
lay the basis) for entirely new approaches for consideration of different
RF-devices. This approach exerted considerable influence on the theory of
slow-wave structures based on utilization of resonant properties of
electromagnetic systems (disk-loaded waveguides, coupled-cavity
chains, etc.)
However, even to this day, there have not been developed
general methods of calculations of small aperture coupling
coefficients from which the  static  results could be obtained by
means of the limit transition $\omega a/c\rightarrow0$ . Development of such
methods would permit not only to assess the region of applicability
of  static  results, but also to expand the frontiers of problems
regarding RF-interactions that can be rigorously solve (correct
evaluation of the separating screen thickness, the vicinity of walls,
etc.) It must be noted that several efforts\footnote{Note, that some
characteristics of RF coupling can be obtained by using variation technique
(see, for example, \cite{r11,r12}). But on the base of this technique
it is difficult to derive the coupled equations that describe the system
under consideration}
were made to push forward
the frontier of applicability of the  static approach in cavity coupling
\cite{r4,r10}. However, the accurateness of the proposed
techniques\footnote{There are plenty of works concerning the diffraction
by a circula conducting disk (or the complimentary problem for a circular
aperture in an infinite plane conducting screen)(see,for example, \cite{r13}),
but the cavity  coupling problem includes the development of the procegure
for deriving specific coupled equations}
cannot be proven
within the framework of the models considered.
Development of novel analytical method for investigation of
electromagnetic system coupling through small-size apertures is
also important considering the fact that there are difficulties of
utilization the widely developed electromagnetic simulations
techniques in this particular area. These difficulties are associated
with the requirements of very high precision mathematical models to
be used for small coupling holes, since the relative correctness of a
model has to be smaller than the coupling coefficients.
This paper presents research results in the development of a
novel analytical model for studies on electromagnetic systems
coupling through small-size apertures (\cite{r14}-\cite{r16}).
Considered is the key problem of two cavity coupling through an aperture
in separating screen of finite thickness without making assumption on
smallness of any parameters.

\section{Problem Definition}

Let us consider two ideal conducting co-axial cylindrical
cavities coupled through a cylindrical aperture of the radius
$ a $   in the separating planar screen of the thickness $ t  $.
The radii and lengths of the first and second
cavities will be designated   $b_1, d_1$ and $b_2, d_2$, respectively.
To construct a mathematical model of the electromagnetic system
under consideration, we will use a relatively novel method of
partial cross-over regions (see, for instance, \cite{r17,r18}). As the first
and second regions, we will take the cylindrical cavity volumes; for the
third, a cylinder that is co-axial with the coupling hole , its radius
being equal $b_3=a$ . This cylinder projects into the area of the
first cavity for the length $d_{1\ast}$  and into the second one for the
length $d_{2\ast}$ , the cylinder length being $l_{\ast}=d_{1\ast}+d_{2\ast}
+t$.

In each region, we expand the electromagnetic fields in terms
of the orthonormal complete set of field functions without the hole:
\begin{eqnarray}
\vec{E}=\sum_{n,s}e_{n,s}^{(i)}\vec{\cal E}_{n,s}^{(i)} +
\sum_{n,s}e_{n,s}^{\prime (i)}\vec{\cal E}_{n,s}^{\prime(i)} \label{qw1}\\
\vec{H}=\sum_{n,s}h_{n,s}^{(i)}\vec{\cal H}_{n,s}^{(i)} +
\sum_{n,s}h_{n,s}^{\prime (i)}\vec{\cal H}_{n,s}^{\prime(i)} \label{qw2}
\end{eqnarray}
where $\vec{\cal E}_{n,s}^{(i)},\vec{\cal H}_{n,s}^{(i)}$  - are solenoidal
and $\vec{\cal E}_{n,s}^{\prime (i)},\vec{\cal H}_{n,s}^{\prime (i)}$  -
irrotational sub- sets. For axial-symmetric modes $e_{n,s}^{\prime
(i)}=h_{n,s}^{\prime (i)}=0$, while the set of solenoidal basic functions
takes on the form:
\begin{eqnarray}
{\cal E}_{n,s,\rm z}^{(i)}=\frac{\lambda_s^2 c}{b_i^2 \omega_{n,s}^{(i)}
N_{n,s}^{(i)}}\cos(k_n^{(i)}\xi_i)J_0\left(\lambda_s r/b_i\right),
\label{qw3} \\
{\cal E}_{n,s,\rm r}^{(i)}=\frac{\lambda_s k_n^{(i)} c}{b_i
\omega_{n,s}^{(i)} N_{n,s}^{(i)}}\sin(k_n^{(i)}\xi_i)J_1\left(\lambda_s
r/b_i\right), \label{qw4} \\
{\cal H}_{n,s,\rm \phi}^{(i)}=-i \frac{\lambda_s }{b_i
N_{n,s}^{(i)}}\cos(k_n^{(i)}\xi_i)J_1\left(\lambda_s
r/b_i\right), \label{qw5}
\end{eqnarray}
where
$$
i=1,2,3; \ s=1,2 \ldots \infty; \ n=0,1 \ldots \infty;
J_0\left(\lambda_s\right)=0;
$$
$$
\omega_{n,s}^{(i)}=c\sqrt{k_n^{(i)2}+\lambda_s^2/b_i^2}; \ k_n^{(i)}=\pi n
/d_{i}; \ N_{n,s}^{(i)}=\sqrt{\pi \theta_n d_i \lambda_s^2
J_1^2(\lambda_s)/2};
$$
$$
\theta_n=\left\{
\begin{array}{lr}
2,&n=0,\\
1,&n\not= 0,
\end{array}
\right. \
\xi_1=z; \ \xi_2=z-d_1-t; \ \xi_3=z-d_1-d_{1\ast}.
$$

 The basic set (\ref{qw3}-\ref{qw5}) satisfies the orthonormality
 condition:
\begin{equation}
\int_v\vec{\cal E}_{n,s}^{(i)}\vec{\cal E}_{n^\prime,s^\prime}^{(i)\ast}dV=
\int_v\vec{\cal H}_{n,s}^{(i)}\vec{\cal H}_{n^\prime,s^\prime}^{(i)\ast}dV=
\delta_{n,n^\prime}\delta_{s,s^\prime}. \label{qw6}
\end{equation}

Coefficients $e_{n,s}^{(i)}$   in the expansion (\ref{qw1}-\ref{qw2}) are
determined by the electric field tangential components at the boundaries of
the chosen regions
\begin{equation}
\left(\omega_{n,s}^{(i)2}-\omega^2\right) e_{n,s}^{(i)}=-ic\omega_{n,s}^{(i)}
\int_{S}\left[\vec{E}\vec{\cal H}_{n,s}^{\ast(i)}\right]d\vec{s}. \label{qw7}
\end{equation}
Since the electric field tangential component $\vec{E_{\tau}}$  on a
metallic surface is zero, then, in Eq.(\ref{qw7}) the integration surfaces
for the first and second regions will be circles located on the opposite
planes of the hole in the screen, while for the third one, two cylindrical
surfaces and two circles, following which this particular region is in
contact with the former two regions. Remembering this, we derive from
(\ref{qw7}) the following:
\begin{equation}
\left(\omega_{k,l}^{(i)2}-\omega^2\right) e_{k,l}^{(i)}=
\sum_{n,s} e_{n,s}^{(3)} L_{n,s,k,l}^{(i)}, \; \;i=1,2, \label{qw8}
\end{equation}
\begin{equation}
e_{n,s}^{(3)}=\sum_{n^\prime,s^\prime}
\left(
e_{n^\prime,s^\prime}^{(1)} T_{n^\prime,s^\prime,n,s}^{(1)}+
e_{n^\prime,s^\prime}^{(2)} T_{n^\prime,s^\prime,n,s}^{(2)}
\right), \label{qw9}
\end{equation}
$$
L_{n,s,k,l}^{(1)}= ic\omega_{k,l}^{(1)}2\pi \int_{0}^{a} rdr{\left({\cal E}_{n,s,\rm
r}^{(3)} {\cal H}_{k,l,\rm{\phi}}^{\ast(1)}\right)}_{z=d_1},
$$
$$
L_{n,s,k,l}^{(2)}= -ic\omega_{k,l}^{(2)}2\pi \int_{0}^{a} rdr{\left({\cal E}_{n,s,\rm
r}^{(3)} {\cal H}_{k,l,\rm{\phi}}^{\ast(2)}\right)}_{z=d_1+t},
$$
$$
T_{n^\prime , s^\prime , n, s}^{(1)}=
\frac{2\pi i c \omega_{n,s}^{(3)}}
{\omega_{n,s}^{(3)2}-\omega^2} \times
$$
$$
\times \left[
-a\int_{d_1-d_{1 \ast}}^{d_1} dz
{
\left( {\cal E}_{n^\prime,s^\prime,\rm z}^{(1)}
{\cal H}_{n,s,\rm{\phi}}^{\ast(3)}\right)
}_{r=a}- \\
\int_0^a rdr
{
\left({\cal E}_{n^\prime,s^\prime,\rm r}^{(1)}
{\cal H}_{n,s,\rm{\phi}}^{\ast(3)}\right)
}_{z=d_1-d_{1 \ast}}
\right],
$$
$$
T_{n^\prime , s^\prime , n, s}^{(2)}=
\frac{2\pi i c \omega_{n,s}^{(3)}}
{\omega_{n,s}^{(3)2}-\omega^2} \times
$$
$$
\times \left[
-a\int_{d_1+t}^{d_1+t+d_{2\ast}} dz
{
\left( {\cal E}_{n^\prime,s^\prime,\rm z}^{(2)}
{\cal H}_{n,s,\rm{\phi}}^{\ast(3)}\right)
}_{r=a}- \\
\int_0^a rdr
{
\left({\cal E}_{n^\prime,s^\prime,\rm r}^{(2)}
{\cal H}_{n,s,\rm{\phi}}^{\ast(3)}\right)
}_{z=d_1+t+d_{2 \ast}}
\right].
$$
Substituting (\ref{qw9}) into (\ref{qw8}), and introducing for the sake of
convenience new variables
$$
a_{k,l}^{(i)}=e_{k,l}^{(i)} \frac{\lambda_l J_0 \left(\lambda_l a/b_i\right)}
{\omega_{k,l}^{(i)} \sqrt{\theta_k} J_1 (\lambda_l) }
$$
instead $e_{k,l}^{(i)} \; (i=1,2)$, we get a set of equations for field
amplitudes only in the 1-st and 2-nd regions
\begin{equation}
\theta_k Z_{k,l}^{(i)} a_{k,l}^{(i)}=
\sum_{n^\prime , s^\prime}
\left(
a_{n^\prime , s^\prime}^{(1)} V_{n^\prime , s^\prime ,k,l}^{(i,1)}+
a_{n^\prime , s^\prime}^{(2)} V_{n^\prime , s^\prime ,k,l}^{(i,2)}
\right),
\; \; i=1,2 \label{qw10}
\end{equation}
where $Z_{k,l}^{(i)}=\omega_{k,l}^{(i)2}-\omega^2$,
$$
V_{n^\prime , s^\prime ,k,l}^{(i,j)}=\sum_{n^\prime , s^\prime}
\frac{
\lambda_l J_1\left(\lambda_{s^\prime}\right)
J_0 \left(\lambda_l a/b_i\right) \omega_{n^\prime, s^\prime}^{(i)}
}
{
\lambda_{s^\prime} J_1\left(\lambda_l\right)
J_0 \left(\lambda_{s^\prime} a/b_i\right) \omega_{n, s}^{(i)}
}
\sqrt{\theta_{n^\prime}\theta_k}
T_{n^\prime , s^\prime , n, s}^{(j)}
L_{n^\prime,s^\prime,k,l}^{(i)}.
$$
After making simple, although cumbersome, calculations we obtain
the following expression for the coefficients
$V_{n^\prime , s^\prime ,k,l}^{(i,j)}$:
$$
V_{n^\prime , s^\prime ,k,l}^{(i,j)}=(-1)^{1+i(1+k)+j(1+n^\prime)}
\alpha_{i,j} \gamma_{l,i} \sum_s \sigma_{s,l,i}\Delta_{s,n^\prime,j}\times
$$
\begin{equation}
\times \left[f_s^{(i,j)}- \beta_i Z_{n^\prime, s^\prime}^{(j)}
\sigma_{s,s^\prime,j} R_{n^\prime,j}F_s^{(i,j)}\right], \label{qw11}
\end{equation}
where
$$
\alpha_{i,j}=4a^3c^2{\left(b_i^2b_j^2\sqrt{d_1d_2}\right)}^{-1},
\gamma_{l,i}=\lambda_l^2J_0^2\left(\lambda_la/b_i\right)/J_1(\lambda_l),
$$
$$
\sigma_{s,l,i}={\left( \lambda_s^2-a^2\lambda_l^2/b_i^2\right) }^{-1},
\Delta_{s,n,j}={\left[ \lambda_s^2- \Omega^2+{\left(\pi a n/d_j \right) }^2
\right] }^{-1},
$$
$$
\beta_i=2a^3/\left( c^2d_i \right), R_{n,j}=\pi n \sin(\pi n d_{j\ast}/d_j),
$$
$$
F_s^{(i,j)}=\frac{1}{\sinh(q_s)}
\left\{
\begin{array}{lr}
\sinh[ q_s \left( 1-d_{i\ast}/l_{\ast} \right) ],& i=j,\\
\sinh[ q_s d_{i\ast}/l_{\ast} ],& i\not=j,
\end{array}
\right.
$$
$$
f_s^{(i,j)}=\frac{\mu_s}{\sinh(q_s)} \times
$$
$$
\times \left\{
\begin{array}{lr}
\cosh[q_s]-\cosh[ q_s \left( 1-2 d_{i\ast}/l_{\ast} \right) ],& i=j,\\
\cosh[q_s \left( d_{2\ast}+d_{1\ast}\right)/l_{\ast)}]-
\cosh[q_s \left( d_{2\ast}-d_{1\ast}\right)/l_{\ast)}],& i\not=j,
\end{array}
\right.
$$
$$
q_s=\mu_s l_{\ast}/a, \; \mu_s=\sqrt{\lambda_s^2-\Omega^2},\; \Omega=\omega
a/c.
$$

The uniform set of Eqs.(\ref{qw10}) describes the interaction of
two infinite sets of oscillators, which are eigenmodes of closed
cavities (without the coupling hole in the separating screen), being,
in principle, fit to be used for calculations of necessary
electromagnetic characteristics of coupled cavities. However, the set
of Eqs.(\ref{qw10}) has three drawbacks that make it difficult
to carry
out both  analytical investigations and  numerical calculations.
Firstly, the structure of this set of equations does not yield a
possibility to obtain analytical results, in particular,  in the well
studded limit  $t=0$ and $a\rightarrow0$. Secondly, this set is
two-dimensional, and it is necessary to have great calculative
resources to solve it. Thirdly, owing to the presence of field singularity
peculiarities at acute angles of the hole in the screen the coefficients
$V_{n^\prime , s^\prime ,k,l}^{(i,j)}$
decrease slowly with increasing indices. Our studies
show that the set of Eqs.(\ref{qw10})  can be reduced to such a form that
has no first or second drawbacks. Below are the results of these
studies.

\section{Derivation of Basic Equations}

Let us seek for the amplitudes of eigenmodes $a_{k,l}^{(i)}$,
except for the fundamental modes ($(k,l)\not=(0,1)$), in the form:
\begin{equation}
\theta_k Z_{k,l}^{(i)} a_{k,l}^{(i)}=a_{0,1}^{(1)} x_{k,l}^{(i,1)}+
a_{0,1}^{(2)} x_{k,l}^{(i,2)}. \label{qw12}
\end{equation}
By introducing two new sequences of unknown values
$\left\{x_{k,l}^{(i,1)}\right\}$ and $\left\{x_{k,l}^{(i,2)}\right\}$,
instead of one $\left\{a_{k,l}^{(i)}\right\}$, we can impose one additional
condition on these new sequences. Let us assume that
$\left\{x_{k,l}^{(i,1)}\right\}$  satisfies the equations
\begin{equation}
x_{k,l}^{(i,1)}={\sum_{n^\prime , s^\prime}}^{\prime}
\left(
\frac{x_{n^\prime , s^\prime}^{(1,1)}}{\theta_{n^\prime}
Z_{n^\prime , s^\prime}^{(1)}} V_{n^\prime , s^\prime ,k,l}^{(i,1)}+
\frac{x_{n^\prime , s^\prime}^{(2,1)}}{\theta_{n^\prime}
Z_{n^\prime , s^\prime}^{(2)}} V_{n^\prime , s^\prime ,k,l}^{(i,2)}
\right)+
V_{0,1,k,l}^{(i,1)}, \label{qw13}
\end{equation}
where $(k,l)\not=(0,1)$, $i=1,2$.

Then from Eqs.(\ref{qw10})  it follows that $\left\{x_{k,l}^{(i,2)}\right\}$
$((k,l)\not=(0,1)$, $i=1,2)$
must satisfy the relationships
\begin{equation}
x_{k,l}^{(i,2)}={\sum_{n^\prime , s^\prime}}^{\prime}
\left(
\frac{x_{n^\prime , s^\prime}^{(1,2)}}{\theta_{n^\prime}
Z_{n^\prime , s^\prime}^{(1)}} V_{n^\prime , s^\prime ,k,l}^{(i,1)}+
\frac{x_{n^\prime , s^\prime}^{(2,2)}}{\theta_{n^\prime}
Z_{n^\prime , s^\prime}^{(2)}} V_{n^\prime , s^\prime ,k,l}^{(i,2)}
\right)+
V_{0,1,k,l}^{(i,2)}, \label{qw14}
\end{equation}
In Eqs.(\ref{qw13},\ref{qw14}) and elsewhere below the prime in sums
indicate that $(n^\prime ,s^\prime)\not=(0,1)$.  It follows from
Eq.(\ref{qw10}) that the amplitudes of fundamental modes
$(k,l)=(0,1)$ should satisfy the
equations
$$
2 Z_{0,1}^{(i)} a_{0,1}^{(i)}=a_{0,1}^{(1)}
\left[
{\sum_{n^\prime , s^\prime}}^{\prime}
\left(
\frac{x_{n^\prime , s^\prime}^{(1,1)}}{\theta_{n^\prime}
Z_{n^\prime , s^\prime}^{(1)}} V_{n^\prime , s^\prime ,k,l}^{(i,1)}+
\frac{x_{n^\prime , s^\prime}^{(2,1)}}{\theta_{n^\prime}
Z_{n^\prime , s^\prime}^{(2)}} V_{n^\prime , s^\prime ,k,l}^{(i,2)}
\right)+
V_{0,1,0,1}^{(i,1)}
\right]+
$$
\begin{equation}
+a_{0,1}^{(2)}
\left[
{\sum_{n^\prime , s^\prime}}^{\prime}
\left(
\frac{x_{n^\prime , s^\prime}^{(1,2)}}{\theta_{n^\prime}
Z_{n^\prime , s^\prime}^{(1)}} V_{n^\prime , s^\prime ,k,l}^{(i,1)}+
\frac{x_{n^\prime , s^\prime}^{(2,2)}}{\theta_{n^\prime}
Z_{n^\prime , s^\prime}^{(2)}} V_{n^\prime , s^\prime ,k,l}^{(i,2)}
\right)+
V_{0,1,0,1}^{(i,2)}
\right]. \label{qw15}
\end{equation}
Let us denote
$$
w_s^{(i,j)}=(-1)^{j+1}3\pi \times
$$
\begin{equation}
\times \left(
f_s^{(i,1)}y_s^{(1,j)}-f_s^{(i,2)}y_s^{(2,j)}-
\beta_1 F_s^{(i,1)}p_s^{(1,j)}+
\beta_2 F_s^{(i,2)}p_s^{(2,j)}
\right), \label{qw16}
\end{equation}
where
$$
y_s^{(i,j)}=\frac{\alpha_{i,i}}{\alpha_{i,j}} \left[
\delta_{i,j} \Delta_{s,0.i}+
{\sum_{n^\prime , s^\prime}}^{\prime}
(-1)^{i\times n^\prime}
\Delta_{s,n^\prime ,i}
\frac{x_{n^\prime , s^\prime}^{(i,j)}}{\theta_{n^\prime}
Z_{n^\prime , s^\prime}^{(i)}}
\right],
$$
$$
p_s^{(i,j)}=\frac{\alpha_{i,i}}{\alpha_{i,j}}
{\sum_{n^\prime , s^\prime}}^{\prime}
(-1)^{i\times n^\prime}
\Delta_{s,n^\prime ,i} \sigma_{s,s^\prime ,i} R_{n^\prime ,s}
\frac{x_{n^\prime , s^\prime}^{(i,j)}}{\theta_{n^\prime} }
$$

Then, Eqs.(\ref{qw15}) can be reduce to the form
\begin{eqnarray}
\left(\omega_{0,1}^{(1)2} -
\omega^2 \right) a_{0,1}^{(1)}=-\omega_{0,1}^{(1)}\frac{2}{3\pi
J_1^2(\lambda_1)} \frac{a^3}{b_1^2 d_1}
\left[a_{0,1}^{(1)}\Lambda_{1,1}-\frac{b_1^2 \sqrt{d_1}}{b_2^2 \sqrt{d_2}}
a_{0,1}^{(2)}\Lambda_{1,2}\right] \label{qw17}\\
\left(\omega_{0,1}^{(2)2} -
\omega^2 \right) a_{0,1}^{(2)}=-\omega_{0,1}^{(2)}\frac{2}{3\pi
J_1^2(\lambda_1)} \frac{a^3}{b_2^2 d_2}
\left[a_{0,1}^{(2)}\Lambda_{2,2}-\frac{b_2^2 \sqrt{d_2}}{b_2^2 \sqrt{d_2}}
a_{0,1}^{(1)}\Lambda_{2,1}\right] \label{qw18}
\end{eqnarray}
where the coefficients $\Lambda_{i,k}$, which define the frequency
shifts and cavities coupling, are determined by the expression
\begin{equation}
\Lambda_{i,k}=J_0\left(\lambda_1 a/b_i \right)
\sum_{s=1}^\infty \sigma_{s,1,i} w_s^{(i,k)}
\label{qw19}
\end{equation}
and   $w_s^{(i,k)}$
are the solutions of the following pair of sets of linear algebraic equations
\begin{equation}
\left\{
\begin{array}{c}
w_m^{(1,1)}+\sum\limits_{s=1}^\infty
\left(
w_s^{(1,1)} G_{m,s}^{(1,1)}+w_s^{(2,1)} G_{m,s}^{(1,2)} \right)
=3\pi f_m^{(1,1)}/\mu_m^2, \\
w_m^{(2,1)}+\sum\limits_{s=1}^\infty
\left(
w_s^{(2,1)} G_{m,s}^{(2,2)}+w_s^{(2,1)} G_{m,s}^{(2,1)} \right)
=3\pi f_m^{(2,1)}/\mu_m^2,
\end{array}
\right.
\label{qw20}
\end{equation}
\begin{equation}
\left\{
\begin{array}{c}
w_m^{(2,2)}+\sum\limits_{s=1}^\infty
\left(
w_s^{(2,2)} G_{m,s}^{(2,2)}+w_s^{(1,2)} G_{m,s}^{(2,1)} \right)
=3\pi f_m^{(2,2)}/\mu_m^2, \\
w_m^{(1,2)}+\sum\limits_{s=1}^\infty
\left(
w_s^{(1,2)} G_{m,s}^{(1,1)}+w_s^{(2,2)} G_{m,s}^{(1,2)} \right)
=3\pi f_m^{(1,2)}/\mu_m^2,
\end{array}
\right.
\label{qw21}
\end{equation}
where
$$
G_{m,s}^{(i,j)}=f_m^{(i,j)} T_{m,s}^{(j)}-
F_{m}^{(i,j)} \delta_{m,s} \frac{\sinh[\mu_m \left(d_i-d_{i\ast}\right)/a]}
{\sinh[\mu_m d_i/a]},
$$
$$
T_{m,s}^{(j)}=
\pi \frac {a} {b} \sum_{s=1}^{\infty}
\frac{ \theta_l^{(j)3} J_0^2(\theta_l^{(j)})
E_l\left(a/d_j,\nu_l^{(j)}\right)} {\chi_l
\left(\lambda_m^2-\theta_l^{(j)2}\right)
\left(\lambda_s^2-\theta_l^{(j)2}\right)}-
\frac {1}{2} \delta_{m,s} E_2\left(a/d_j,\mu_m\right)+
$$
$$
+\frac{ \pi a^2 }{\mu_m^2 b_j d_j}
\frac{\theta_1^{(j)3} J_0^2\left(\theta_1^{(j)}\right)}
{\left(\lambda_m^2-\theta_1^{(j)2}\right)
\left(\lambda_s^2-\theta_1^{(j)2}\right)},
$$
$$
E_l(x,y)=\left\{
\begin{array}{lr}
\coth(y/x)/y-x/y^2,& \ \l=1, \\
\coth(y/x)/y,& \ l \neq 1,
\end{array}
\right.
$$
$$
\theta_l^{(j)}=a\lambda_l/b_j, \;
\chi_l=\pi\lambda_l J_1^2(\lambda_l)/2,\; \nu_l^{(j)}=\sqrt{\theta_l^{(j)2}-
\Omega^2}.
$$

Thus, the set of Eqs.(\ref{qw10}), describing the coupling of
infinite number of oscillators (eigenmodes of closed cavities), has
been rigorously reduced to such a form that formally describes the
interaction of two basic oscillators. In the case considered,
$E_{010}$  modes of the closed cavities were chosen to be such
basic oscillators. Yet, this choosing is arbitrary and determined as the
problem requirements demand. The entire spectrum of the resonance
properties of the coupled cavity system according to this approach is
contained in the dependence of $\Lambda_{i,k}$  on frequency. Such form
of description of the coupled cavities is convenient for solving many
such problems in which electromagnetic characteristics are studied
in a limited frequency range determined by the interactions of two
adjacent eigenmodes.

\section{Research Results}

Before starting to discuss the results of analysis of
Eqs.(\ref{qw17},\ref{qw18}), let us dwell on the problem of choosing the
geometrical dimensions of the auxiliary region 3, namely on the choice of
$d_{i\ast}$
values.  Results of our calculations show that the solution of the linear
algebraic equations obtained by the truncation of infinite systems
(\ref{qw20},\ref{qw21}) at appropriate $S$  and $L$ values do not depend on
$d_{i\ast}$
(we designate the maximum value of the index $s$ in (\ref{qw20},\ref{qw21})
as $S$  and the maximum
value of the index $l$ in the sum that determine $T_{m,s}^{(j)}$  as $L$).
Thus, for instance, at $S=100$, $L=40000$, $d_1=d_2=3.5$~cm, $b_1=b_2=4$~cm,
$t=0.4$~cm, $a=1$~cm, $f=0$ and $d_{\ast}=d_{1 \ast}=d_{2 \ast}$ the
calculations yield: $d_{\ast}=3.5$~cm --- $\Lambda_{1,1}=0.773125$,
$d_{\ast}=10^{-7}$~cm --- $\Lambda_{1,1}=0.773125$.

Let us consider the
case of small-size apertures $(a\rightarrow0)$ and an infinitely thin screen
$(t=0)$
that has been well studied to (\cite{r1}-\cite{r7}). As follows from
Eqs.(\ref{qw17},\ref{qw18}), to compute the values of coupling coefficients
with an accuracy of up to $a^3$ the calculations of coefficients
$\Lambda_{i,k}$ must be performed in the approximation $a=0$. A question
arises of how one should calculate coefficients $T_{m,s}^{(j)}$, since each
term of the sum that determines these coefficients tends to zero at
$a\rightarrow0$. If
this sum converges to any non-zero values, than one must take into account
the infinite numbers of addends. This circumstance reflects the presence of
singularity of the electromagnetic field on the aperture edges.  This fact
makes it practically impossible to employ the initial equations set
(\ref{qw10}) for calculating the characteristics of the coupled cavities in
the case of small apertures. The modified set of equations
(\ref{qw17},\ref{qw18}) can be used to obtain solutions at small $a$-values
with an accuracy of $a^3$. In the limit $a\rightarrow0$ the contribution of
addends with small $l$ in the value of the sum, which determines $T_{m,s}^{(j)}$,
tends to
zero, while at large $l$, the difference of the adjacent Bessel function
roots tends to the value $\pi$ $(\lambda_{l+1}-\lambda_l \approx 0)$, and the
above sum tends to an integral independent of geometrical parameters
$$
\lim_{a\rightarrow0}\pi \frac{a}{b_j} \sum \limits_{l=1}^\infty
\frac{ \theta_l^{(j)3} J_0^2(\theta_l^{(j)})
E_l\left(a/d_j,\nu_l^{(j)}\right)}
{\chi_l \left(\lambda_m^2-\theta_l^{(j)2}\right)
\left(\lambda_s^2-\theta_l^{(j)2}\right)}=
$$
\begin{equation}
=\int_0^\infty
\frac{ J_0^2(\theta) \theta^{2} d\theta}
{\left(\lambda_m^2-\theta^{2}\right)
\left(\lambda_s^2-\theta^{2}\right)}=K_{m,s}. \label{qw22}
\end{equation}

Since in the limit considered $f_m^{(i,j)}\rightarrow\lambda_m$,
then from
(\ref{qw20},\ref{qw21}) it follows that $w_m^{(i,j)}\rightarrow w_m$, where
$w_m$ is the solution of the equation set
\begin{equation}
\sum \limits_{s=1}^\infty K_{m,s} w_s= 3\pi /\left(2 \lambda_m^2\right).
\label{qw23}
\end{equation}
Coefficients $\Lambda_{i,k}$  do not depend on geometrical
dimensions of cavities $\Lambda_{i,k}=\Lambda = \sum_s w_s/\lambda_s^2$.
We have obtained
the value of the constant $\Lambda$ analytically: \, $\Lambda=1$.
Numerical calculations of the truncated
set (\ref{qw20},\ref{qw21}) also showed that with increasing $S$ $\Lambda$
tends to $1$. This is in good agreement with the results of other authors
(see \cite{r1}-\cite{r7}).
Thus, at $S=200$ and an appropriate choice of the step and the integrate
interval in (\ref{qw22}), when the solution of the set (\ref{qw23}) becomes
independent of these parameters, $\Lambda=0,9989$.

As different from earlier
approaches (see, for example, \cite{r1} - \cite{r4}), our equations set
(\ref{qw20},\ref{qw21})
permits to calculate the dependence of electromagnetic characteristics on any
parameters, because no assumptions were made while obtaining it. On the base
of such set one can easily calculate eigenfrequencies. We shall not dwell on
it, but we shall perform analysis of the dependence of coefficients on such
parameters that were impossible to study in earlier models.

First of all,
let us become clear on the influence of electromagnetic field
non-potentiality in the interaction region on $\Lambda_{i,k}$-values, since
in all previous research studies (\cite{r1}-\cite{r4}, \cite{r7}) on
coupling through small-size holes the assumption about field potentiality in
the vicinity of the hole were made. In our model investigation of this
problem is reduced to studying the dependence of the coefficients
$\Lambda_{i,k}$  on the
frequency $f$; the case $f=0$ corresponds to the assumption of field
potentiality in the interaction region.

Since frequency comes into the
appropriate coefficients only in the form of expression $\Omega = \omega a/c$,
then it follows that $\Lambda$-variation with increasing frequencies
from $0$ to $f_{010}$ must be dependent on coupling aperture size - the
smaller $a$ the weaker dependence of $\Lambda_{i,k}$ on frequency. This is
confirmed by the calculations results (Tab.\ref{tb1}).

From Tab.\ref{tb1} it follows that an error in calculations of the coupling
coefficients $\tilde{\Lambda}$ $\left(\tilde{\Lambda}=
2 a^3 \Lambda / \left( 3\pi b_1^2 d_1 J_1^2 \left( \lambda_1 \right)
\right)\right)$ at\footnote{Dimensions of the coupling hole of widely
used disc-loaded waveguides at the operating frequency $f\approx3$\,GHz are
$a=0.9\div 1.5$\,cm.}  $a=1$\,cm is on the order of $10^{-4}$
(the equivalent frequency shift being $\approx300$\,kHz) and,
consequently, all calculations can be  made in static approximation.
Yet, already for $a=1.5$\,cm the error is of the order of $2\,10^{-3}$
(the equivalent frequency shift being $\approx 6$\,MHz), which is
inadmissible for precise calculations.

\begin{table}
\caption{
Dependence
of $\Lambda_{i,k}=\Lambda$  coefficients
and coupling
coefficients $\tilde{\Lambda}$  on frequency $f$
($d_1=d_2= 3.5$\,cm, $b_1=b_2=4$\,cm, $t=0.0$\,cm,
$f_{010}=2.868563$\,GHz)
}
\label{tb1}
\begin{center}
\begin{tabular}{|c|c|c|c|c|} \hline
\multicolumn{1}{|c|}{}&\multicolumn{2}{|c|}{$a=1$\,cm}&
\multicolumn{2}{|c|}{$a=1.5$\,cm} \\   \hline
\raisebox{-.5 ex}{f \, (GHz)} &
\raisebox{-.5 ex}{$\Lambda$} &
\raisebox{-.5 ex}{$\tilde{\Lambda}$} &
\raisebox{-.5 ex}{$\Lambda$} &
\raisebox{-.5 ex}{$\tilde{\Lambda}$} \\ \hline
0 & 0.896590 & 0.012606 & 0.788984 & 0.037440 \\ \hline
1 & 0.897783 & 0.012623 & 0.793784 & 0.037667 \\ \hline
2 & 0.900862 & 0.012666 & 0.808207 & 0.038352 \\ \hline
3 & 0.903614 & 0.012705 & 0.831250 & 0.039445 \\ \hline
\end{tabular}
\end{center}
\end{table}

Of importance for
applied use is the dependence of the coupled coefficients on the coupling
aperture radius $a$.  Analysis of the expression (\ref{qw19}) indicates that
$\Lambda_{i,k}$
depends both on the above parameter $\Omega = \omega a/c$ and on relation of
$a$ to all cavity geometrical parameters and screen thickness ($a/d_j , a
/b_j , a/t , j=1,2$).  Results of the calculations of the relationship of
interest on basic of our model in the static approach ($f =0$)
are given in Tab.\ref{tb2}.

\begin{table}
\caption{
Dependence
of $\Lambda_{i,k}$  coefficients
on the coupling aperture radius $a$
($d_1=d_2= 3.5$\,cm, $b_1=b_2=4$\,cm, $f=0$) }
\label{tb2}
\begin{center}
\begin{tabular}{|c|c|c|c|c|c|} \hline
\multicolumn{1}{|c|}{}&
\multicolumn{1}{|c|}{$\Lambda_{i,k}=\Lambda $}&
\multicolumn{2}{|c|}{$\Lambda_{1,1} $}&
\multicolumn{2}{|c|}{$\Lambda_{1,2} $} \\
\multicolumn{1}{|c|}{a,\,cm}&
\multicolumn{1}{|c|}{$(t=0)$}&
\multicolumn{2}{|c|}{$(t=0.4)$\,cm}&
\multicolumn{2}{|c|}{$(t=0.4)$\,cm} \\   \hline
\multicolumn{1}{|c|}{}&
\multicolumn{1}{|c|}{}&
\multicolumn{1}{|c|}{}&
\multicolumn{1}{|c|}{\cite{r9}}&
\multicolumn{1}{|c|}{}&
\multicolumn{1}{|c|}{\cite{r9}} \\ \hline
0.04 & 0.9976 & 0.8580 & 0.8584 & 0.0000 & 0.0000 \\ \hline
0.10 & 0.9965 & 0.8571 & & 0.0000  & \\ \hline
0.133333 & 0.9956 & 0.8563 & 0.8584 & 0.0006 & 0.0006 \\ \hline
0.20 & 0.9929 & 0.8539 & & 0.0066  & \\ \hline
0.30 & 0.9871 & 0.8489 & & 0.0331   & \\ \hline
0.40 & 0.9792 & 0.8422 & 0.8590 & 0.0734 & 0.0748 \\ \hline
0.50 & 0.9695 & 0.8341 & & 0.1181    & \\ \hline
0.70 & 0.9448 & 0.8138 & & 0.2016   &  \\ \hline
0.90 & 0.9140 & 0.7880 & & 0.2675   &  \\ \hline
1.00 & 0.8965 & 0.7731 & & 0.2934   &  \\ \hline
1.10 & 0.8777 & 0.7568 & & 0.3150   &  \\ \hline
1.333333 & 0.8286 & 0.7139 & 0.8765 & 0.3500 & 0.4202 \\ \hline
1.30 & 0.8360 & 0.7204 & & 0.3462   & \\ \hline
1.50 & 0.7889 & 0.6790 & & 0.3633   & \\ \hline
\end{tabular}
\end{center}
\end{table}

Tab.\ref{tb2} also shows the results of calculations for various values of
the parameter $a/t$ at $a/d_j\rightarrow 0, a/d_j\rightarrow 0$ \cite{r9}.
It follows from Tab.\ref{tb2} that taking into account the finitness of
parameters  $a/d$ and $a/b$  lead not only to a drastic change of the
numerical values of the coupling coefficients, but to change the functional
dependence of $\Lambda_{i,k}$  on $a$.  For instance, at a finite thickness
of the screen the coefficient $\Lambda_{1,1}$, which determines the cavity
eigenfrequency shift, decreases with increasing $a$, contrary to what one
can obtain from the results of the paper \cite{r9}.

From the basic set of Eqs.(\ref{qw17},\ref{qw18}) it follows that for
identical cavities the dependence of the coupling coefficients on the cavity
length $d$  is determined by the ratio
 $\Lambda_{i,k}/d$. The presently developed models of coupled
cavity model (\cite{r1}-\cite{r7}) are true when the condition
$a /d \ll 1$ is satisfied. In this case $\Lambda_{i,k}$  are
independent of $d$  and coupling coefficients are inversely
proportional to $d$.
\begin{table}
\caption{
Dependence
of coefficients $\Lambda_{i,k}/d$  on the cavity lenght $d_1=d_2=d$
($\Lambda_{i,k}=\Lambda$ --- $t=0$, $\Lambda_{1,1}, \Lambda_{1,2}$ ---
$t=0.4$\,cm,
$b_1=b_2=4$\,cm,
$f=0$)
}
\label{tb3}
\begin{center}
\begin{tabular}{|c|c|c|c|} \hline
\multicolumn{1}{|c|}{$d$,\,cm}&
\multicolumn{1}{|c|}{$\Lambda_{i,k}/d=\Lambda/d$}&
\multicolumn{1}{|c|}{$\Lambda_{1,1}/d$}&
\multicolumn{1}{|c|}{$\Lambda_{1,2}/d$}  \\
\multicolumn{1}{|c|}{}&
\multicolumn{1}{|c|}{$(t=0)$}&
\multicolumn{1}{|c|}{$(t=0.4)$\,cm}&
\multicolumn{1}{|c|}{$(t=0.4)$\,cm} \\   \hline
0.001 &  2.559185  &      5.071967  &      0.010858    \\   \hline
0.050 &  2.434071  &      3.851450  &      0.357559    \\   \hline
0.100 &  2.305363  &      3.198808  &      0.512277    \\   \hline
0.200 &  2.062168  &      2.425438  &      0.598867    \\   \hline
0.300 &  1.839241  &      1.952942  &      0.579913    \\   \hline
0.500 &  1.460225  &      1.383953  &      0.478433    \\   \hline
0.750 &  1.112953  &      0.992879  &      0.365433    \\   \hline
1.000 &  0.877394  &      0.764395  &      0.287444    \\   \hline
1.500 &  0.600848  &      0.516681  &      0.196466    \\   \hline
2.000 &  0.451742  &      0.388012  &      0.147701    \\   \hline
2.500 &  0.360623  &      0.310091  &      0.117952    \\   \hline
3.000 &  0.299667  &      0.258049  &      0.098056    \\   \hline
3.500 &  0.256169  &      0.220893  &      0.083854    \\   \hline
\end{tabular}
\end{center}
\end{table}

Results of our calculations, as one can see from Tab.\ref{tb3},
indicate that in the
region  $a/d > 1$ the dependence of coupling
coefficients on $d$  changes significantly; namely, at
$d\rightarrow 0$ $\Lambda_{i,k}/d$  tend to constant values that
are determined by other cavity geometrical dimensions. For instance,
for the case  $t\not= 0$ at  $d\rightarrow 0$ the coefficient
$\Lambda_{1,2} /d$, which determined the frequency difference
between $0$ and $\pi$-type eigenmodes of the system, tends to zero,
while the coefficient $\Lambda_{1,1} /d$, determining the
eigenfrequency shift, tends to non-zero
value. This signifies that in the two-thin-cavity case ($d/t\ll 1$)
difference between $0$ and $\pi$ type frequencies will be small. This result
is in agreement with the circumstance that at $d=0$  the two
cavity system transforms into a single cavity with the radius
$r=a$ and the length $t$. However, in this case one
must takes into account the dependence of parameters on frequency.

\section{Conclusion}

On this way, we put forward a novel analytical model for
studding the coupling of two cavities through an aperture in
separating screen of finite thickness without making assumption
on smallness of any parameters.

1. On the base of rigorous electromagnetic
approach the coupling coefficients of the cylindrical cavities
within the limit of small aperture and infinitely thin wall are
calculated.

2. The presented numeric results of electromagnetic
characteristic dependencies that have been impossible to perform
on the base of previous models show that influence of non-potentiality
of fields in the vicinity of the hole and nearness of
cavity walls may be considerable.


\begin{thebibliography}{99}
\bibitem{r1} H.A.~Bethe. Phys. Rev., 1944, v.66, N.7, p.163-182.
\bibitem{r2} V.V.~Vladimirsky. ZhTF, 1947, v.17, N.11, p.1277-1282.
\bibitem{r3} A.I.~Akhiezer, Ya.B. Fainberg. UFN, 1951, v.44, N.3, p.321-368.
\bibitem{r4} R.M.~Bevensee. Electromagnetic Slow Wave Systems. John
Wiley\&Sons, Inc.,New York-London-Sydney, 1964.
\bibitem{r5} H.A.~Wheeler. IEEE Trans. on Microwave Theory and
Techniques, 1964, MTT-12, p.231-244.
\bibitem{r6} R.L.~Gluckstern, R. Li, R.K. Copper. IEEE Trans. on Microwave
Theory and Techniques, 1990, MTT-38, N.2, p.186-192.
\bibitem{r7} R.L.~Gluckstern. AIP Conference Proceedings 249, v.1, The
Physics of Particle Accelerators, AIP, New York, 1992, p.236-276.
\bibitem{r8} N.A.~McDonald. IEEE Trans. on Microwave Theory and
Techniques, 1972, MTT-20, N.10, p.689-695.
\bibitem{r9} R.L.~Gluckstern, J.A. Diamond. IEEE Trans. Microwave
Theory and Techniques, 1991, MTT-39, N.2, p.274-279.
\bibitem{r10} N.P.~Sobenin, E.Ya. Shkolnicov. Collected Series:
Accelerators, Moscow, Atomizdat, 1970, N.12, p.96-101.
\bibitem{r11} W-H.~Cheng, A.V.~Fedotov, R.L.~Gluckstern.
Phys.Rev.E, 1995, v.E52, N3, p.3127-3142.
\bibitem{r12}  McDonald~N.A. IEEE Trans. Microwave Theory Tech., 1972,
v.MTT-20, N10, p.689-695.
\bibitem{r13} C.M.~Butler, Y.~Rahmat-Samii, R.~Mitra.
IEEE Trans. on Antennas and Propagation., 1978, v.AP-26, N.1, p.82-93.
\bibitem{r14} M.I.~Ayzatsky. On two-cavity coupling. Preprint NSC KPTI 95-8,
1995.
\bibitem{r15} M.I.~Ayzatsky. Proc.14th Workshop on Charged Particle
Accelerators. Protvino, 1994, vol.1, p.240.
\bibitem{r16} M.I.~Ayzatsky. ZhTF. 1996, vol.66, in publication.
\bibitem{r17} I.G.~Prohoda, V.I.~Lozyanoi, V.M.~Onufrienko et al.
Electromagnetic wave propagation in inhomogeneous waveguide systems.
Dnepropetrovsk, Dnepropetrovsk State University Publishing House, 1977.
\bibitem{r18} I.G.~Prohoda, V.P.Chumachenko. Izvestia VUZov, Radiophisics,
1993, v.16, N.10, p.1558.
\end{thebibliography}
\end{document}